\documentclass{iau} 
\usepackage{graphicx}
\usepackage{hyperref}

\title[HASH PN Database] 
{The Hong Kong/AAO/Strasbourg H$\alpha$ (HASH) Planetary Nebula Database}

\author[I. S. Boji\v{c}i\'c, Q. A.~Parker \& D. J. Frew]   
{Ivan S. Boji\v{c}i\'c$^{1,2}$, Quentin A.~Parker$^{1,2}$
 \and David J. Frew$^{1,2}$}

\affiliation{$^1$1 Department of Physics, The University of Hong Kong\\
Hong Kong SAR, China\\ 
[\affilskip]
$^2$Laboratory for Space Research, The University of Hong Kong\\
Hong Kong SAR, China\\
email: {\tt ibojicic@hku.hk}}

\pubyear{2017}
\volume{323}  
\jname{Planetary Nebulae: Multi-Wavelength Probes of Stellar and Galactic Evolution}
\editors{A.C. Editor, B.D. Editor \& C.E. Editor, eds.}
\begin{document}

\maketitle

\begin{abstract}
The Hong Kong/AAO/Strasbourg H$\alpha$ (HASH) planetary nebula database is an online research platform providing free and easy access to the largest and most comprehensive catalogue of known Galactic PNe and a repository of observational data (imaging and spectroscopy) for these and related astronomical objects. The main motivation for creating this system is resolving some of long standing problems in the field e.g. problems with mimics and dubious and/or misidentifications, errors in observational data and consolidation of the widely scattered data-sets. This facility allows researchers quick and easy access to the archived and new observational data and creating and sharing of non-redundant PN samples and catalogues.

\end{abstract}

\firstsection 

\section{Architecture}

\begin{figure}[t]
\begin{center}
\includegraphics[height=1.7in]{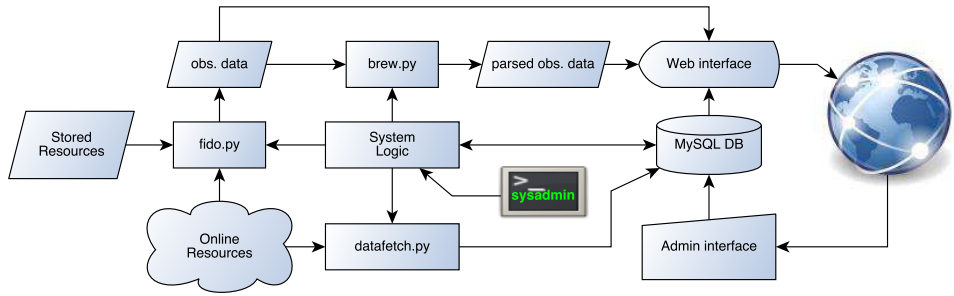}
 \caption{HASH system flowchart. After a new object is added to the database its corresponding imaging and tabulated data are first automatically collected from online and local resources using in-house written python scripts. The initial dataset is then manually vetted by the database administrators (the co-authors) applying position corrections, correcting morphological class etc, and then released to the publicly accessible web interface. We also encourage users to suggest corrections or modifications through the system itself (by adding a note on the object's individual page). }
 \label{fig:flowchart}
\end{center}
\end{figure}

The HASH PN database (HASH) and its online interface (\url{http://hashpn.space}) is currently hosted and maintained on a dedicated server located on The University of Hong Kong network. The main technologies used in the web interface are PHP and JavaScript. The backend of the system (i.e. image/spectra/data parsers, image/spectra visualisation and data analysis) consist of in-house scripts written in python using several freely available scientific computing, astronomical and plotting modules (e.g. APLpy, astropy)\footnote{http://www.astropy.org/; https://aplpy.github.io/}. The basic full system (frontend and backend) flowchart is presented in Fig. \ref{fig:flowchart} and is similar to other astronomical databases (e.g. Moreau \& Dubernet 2006). 

The backbone of HASH is the relational database (MySQL) which provides data consistency and quick, efficient search and manipulation. Each HASH object is characterised by a unique set of parameters (i.e. unique id, coordinates of the nebular centroid and the PN~G designation) and mapped to the corresponding observational data collected from the literature and astronomical data repositories.

\section{HASH Content}

Over 6700 Galactic objects are currently in our working database, of which $\sim$3500 are designated as true, likely and possible PNe (for more details and the base catalogue construction and PN identification methods see \cite{Parker2016}, \cite[Frew \& Parker, 2010]{Frew2010} and Parker \etal, these proceedings). The observational data repository contains over 300k fits image cutouts from over 30 large scale imaging surveys. We mirrored the full data from some of these surveys on our server (e.g. SHASSA, NVSS, etc) thus allowing the benefit of fast data access. On the other hand, high resolution, ongoing surveys and large datasets (e.g. 2MASS, UKIDSS) are queried on demand using provided data access services or, preferably, Simple Image Access Protocol (SIAP; for details see IVOA Recommendation documents at \url{http://www.ivoa.net/documents/}). HASH contains over 7000 reduced spectra and over 85000 measured spectral line fluxes (relative and/or flux calibrated). One of the most important emission line data sources is the Electronic Emission-Line Catalog for Planetary Nebulae (ELCAT; \cite[Kaler et al., 1997]{Kaler1997}). We extended ELCAT  with data 
for $\sim$ 500 Galactic PNe published after ELCAT was first released. 

The tabulated data for objects in the database is parsed from Vizier or directly from the literature. In this version we did not include any derived parameters found in the literature (distances, physical sizes, metalicities, etc.) mainly because of inhomogeneities in methods and data. 

In addition to already published observational data we provide new and more accurate informations on PN positions, angular diameters (in optical bands) and morphologies. Astrometric positions reported here are best estimates of the position of the centroid of the nebula. Positions of ionising (central) stars are catalogued separately in order to enable scrutiny and analysis of that data set separately.

\section{Future work}

The full catalogue of known Galactic PNe and its accompanying observational data will soon be published and integrated into the CDS database (Boji\v{c}i\'{c} \etal\ in preparation). As we continue to detect new PNe and confirm additional PN candidates, together with the prospect of more accurate distance estimates (\cite[Frew et al. 2016]{Frew2016}, Stanghellini et al. these proceedings) the next version of HASH will contain a more complete sample of Galactic PNe as well as including a large range of homogeneously derived physical parameters.


\begin{thebibliography}{}

\bibitem[Frew \& Parker(2010)]{Frew2010}
{Frew, D.~J., \& Parker, Q.~A.,} 2010,
\textit{PASA}, 27,129 

\bibitem[Frew et al.(2016)]{Frew2016} 
{Frew, D.~J., Parker, Q.~A., \& Boji{\v c}i{\'c}, I.~S.,} 2016, 
\textit{MNRAS}, 455, 1459 

\bibitem[Kaler et al.(1997)]{Kaler1997}
{Kaler, J.~B., Shaw, R.~A., \& Browning, L.,} 1997,
\textit{PASP}, 109, 289

\bibitem[Moreau \& Dubernet]{MD06}
{Moreau, N., \& Dubernet, M.~L.,} 2006, \textit{ASP Conf.Ser.}, 351, 391

\bibitem[Parker \etal\ (2016)]{Parker2016}
{Parker, Q.~A., Boji{\v c}i{\'c}, I.~S., \& Frew, D.~J.,} 2016, \textit{JPhCs},  728, 032008 


\end{thebibliography}
\end{document}